\documentclass[a4paper, american]{article} 

\usepackage[utf8]{inputenc}
\usepackage{babel}
\usepackage[babel]{microtype}

\usepackage[textwidth=16cm, textheight=23cm]{geometry}

\usepackage{multirow,bigstrut}
\usepackage{amsmath}
\usepackage{amssymb}
\usepackage{bm}
\usepackage{dsfont}
\usepackage{authblk}
\usepackage{placeins}
\usepackage[toc,page,titletoc]{appendix}

\usepackage{soul}
\usepackage[dvipsnames]{xcolor}
\usepackage{graphicx}
\usepackage{tikz}
\usepackage{pgfplots}
\usepackage{xfrac}
\usepackage{mathrsfs,amsmath}
\usepackage{siunitx} 
\usepackage{makecell}
\usepackage{comment}
\usepackage{booktabs}

\usepackage{caption}
\usepackage{subcaption}

\usepackage[babel]{csquotes}

\usepackage[numbers]{natbib} 
\usepackage{hyperref}
\hypersetup{
	linkbordercolor=green,
	pdftitle={SBS_amplifier
	},
	pdfauthor={Hui Cao, Doug Stone, Stefan Rothe, Joe Chen, Peyman Ahmadi, Kabish Wisal, Mert Ercan},
	bookmarksnumbered=true,
}

\definecolor{plot0}{HTML}{1bb34c}
\definecolor{plot1}{HTML}{242bb3}
\definecolor{plot2}{HTML}{b3120e}
\definecolor{plot3}{HTML}{cc9f2b}
\definecolor{plot4}{HTML}{27AFB3}
\definecolor{plot0}{HTML}{004488}
\definecolor{plot1}{HTML}{DDAA33}
\definecolor{plot2}{HTML}{BB5566}
\definecolor{plot3}{HTML}{000000}
\definecolor{plot4}{HTML}{AAAAAA}

\pgfplotscreateplotcyclelist{lineplot cycle}{%
{plot0, thick, mark=*},
{plot1, thick, mark=square*},
{plot2, thick, mark=triangle*, mark size=2.3pt},
{plot3, thick, mark=diamond*},
}

\author[1,5]{Stefan Rothe}
\author[1,2,5]{Chun-Wei Chen}
\author[3,1,5]{Peyman Ahmadi}
\author[4]{Kabish Wisal}
\author[1]{Mert Ercan}
\author[1]{KyeoReh Lee}
\author[1]{Nathan Vigne}
\author[1]{A. Douglas Stone}
\author[*,1]{Hui Cao}

\affil[1]{Department of Applied Physics, Yale University, New Haven, CT 06520, USA}
\affil[2]{Edward L. Ginzton Laboratory, Stanford University, Stanford, CA 94305, USA}
\affil[3]{Coherent, 1280 Blue Hills Ave., Bloomfield, 06002, CT, USA}
\affil[4]{Department of Physics, Yale University, New Haven, CT 06520, USA}
\affil[5]{These authors contributed equally: Stefan Rothe, Chun-Wei Chen, Peyman Ahmadi}
\affil[*]{Corresponding author: hui.cao@yale.edu}
\date{}                     
\title{

Wavefront shaping enables high-power multimode fiber amplifier with output control

}

\begin{document}
	\maketitle
	
	\section*{Abstract}
	
     Over the past two decades there have been tremendous advances in high-power fiber lasers, which have provided a powerful tool for science, engineering and defense. A major roadblock for further power scaling of single-frequency fiber laser amplifiers is stimulated Brillouin scattering. Intense efforts were devoted to mitigate this nonlinear process, but mostly limited to single-mode or few-mode fiber amplifiers which have good beam quality. Here we explore a highly multimode fiber amplifier, where stimulated Brillouin scattering is greatly suppressed due to reduction of light intensity in a large fiber core and broadening of Brillouin scattering spectrum by multimode excitation. To control the output beam profile, we apply spatial wavefront shaping technique to the input light of a nonlinear amplifier to focus the output beam to a diffraction-limited spot outside the fiber facet. Our multimode fiber amplifier can operate at high power with high efficiency and narrow linewidth which ensures high coherence. Optical wavefront shaping enables coherent control of multimode laser amplification, with potential applications in coherent beam combining, large-scale interferometry and directed energy delivery.
    
	
	
	\section{Introduction}
	
	High-power lasers enabled a wide range of applications in research, technology, and medicine~\cite{richardson2010high, nilsson2011high, Cesar2013high, Zervas2013high}. The ultimate limit on power scaling of such lasers are nonlinear effects and/or material damage encountered during light amplification inside lasers. An important technique to overcome such detrimental effects in short-pulse amplifiers is chirped pulse amplification~(CPA): laser pulses are stretched temporally, then amplified, and finally compressed again~\cite{strickland1985compression}. A similar technique for continuous-wave~(CW) amplifiers is spatial spread of light, but it is difficult to implement in optical-fiber based amplifiers that typically have a small cross section. Increasing the fiber core area will likely introduce additional guided modes, and the interference of light in these modes will generate a pseudo-random spatial field distribution, leading to poor output beam quality. Since high beam quality is needed for many applications like laser machining, interferometry and directed energy delivery, most narrowband fiber amplifiers operate with single or quasi-single spatial modes, but strong light confinement in small-core fibers makes suppression of optical nonlinearity and instability much more challenging. 
	
	The lowest-power nonlinear limit in power-scaling of narrowband fiber amplifiers typically arises from stimulated Brillouin scattering~(SBS) of light mediated by acoustic waves~\cite{kobyakov2010stimulated, panbhiharwala2018investigation}. This effect scatters forward propagating signal to backward Stokes light, limiting the amplifier output power. Moreover, intense Stokes pulses might be generated and they could damage upstream lasers. Various techniques have been developed to mitigate SBS, mostly in the single-mode or few-mode fibers, to maintain high output-beam quality. One commonly used method to increase SBS threshold is increasing the signal linewidth to tens of GHz to effectively broaden the Brillouin scattering spectrum and thereby lowering the peak gain for SBS~\cite{flores2014pseudo}. However, this approach greatly reduces temporal coherence that is required for applications like coherent beam combining and gravitational-wave detection. 
	To suppress SBS in the single-frequency (linewidth < 100 kHz) regime, different approaches have been explored~\cite{yoshizawa1993stimulated, kobyakov2005design, dragic2005optical, li2007ge, liu2007suppressing, robin2011acoustically, limpert2012yb, robin2014modal, hawkins2021kilowatt, jiang2022650, shi2022700}, but SBS remains the major obstacle for further power scaling~\cite{li2023high}.

	In recent years, optical wavefront shaping has become a powerful technique for controlling light propagation in complex media~\cite{mosk2012controlling, rotter2017light} including multimode fibers~(MMFs)~\cite{cao2022shaping}. For example, tailoring the incident wavefront of light to an MMF with a spatial light modulator~(SLM) can create any spatial pattern of transmitted field that is a superposition of the fiber modes~\cite{di2010hologram, papadopoulos2012focusing, choi2012scanner, caravaca2013real, stellinga2021time, gomes2022near, michalkova2024generating}. Most of these studies were conducted on passive MMFs with linear light propagation or low-power MMF amplifiers with weak nonlinearity~\cite{florentin2017shaping, florentin2018space, florentin2019shaping, wei2020harnessing}. Adaptive wavefront shaping has been employed to control nonlinear optical processes in passive MMFs~\cite{tzang2018adaptive, deliancourt2019wavefront, wright2022nonlinear}, but it has not been applied to high-power MMF amplifiers with strong nonlinearity, and its ability, efficacy and robustness of controlling complex, unstable multimode amplification are not yet known.
	
	Here we simultaneously suppress detrimental SBS and tailor output beam profile in a highly multimode nonlinear fiber amplifier with input wavefront shaping. In a large-core MMF, transverse spread of light lowers intensity and reduces SBS. Additionally, distributing light among many fiber modes further suppresses SBS, as multimode excitation broadens the SBS spectrum and slows down the Stokes growth \cite{wisal2024theorySBS}. Because of these two factors, our MMF amplifier is free of SBS up to 503~W output power, which is about 30 times higher than the SBS limit in a standard single-mode fiber~(SMF) of same length. The slope efficiency is 82\%, comparable to SMF amplifiers. The output of our MMF amplifier has a 20-dB linewidth of $18$~kHz, corresponding to a 3-dB linewidth of $1$~kHz. This linewidth is 6 orders of magnitude narrower than that of spectrally broadened SMF amplifiers, leading to 6 orders of magnitude longer temporal coherence length.  
	
	To address the main concern of output beam quality for MMF amplifier, we focus the output beam to a diffraction-limited spot by optimizing the input phase front with a SLM. The beam propagation factor for the focal spot is $M^2 \leq 1.35$. Instead of placing the SLM at the amplifier output end, our scheme of shaping the low-power seed with the SLM avoids high-power handling. The ability of  controlling output beam shape, together with high temporal coherence of our MMF amplifier, will greatly facilitate coherent combining of output beams from multiple amplifiers.

	\begin{figure}
		\centering
		\includegraphics[width=\textwidth]{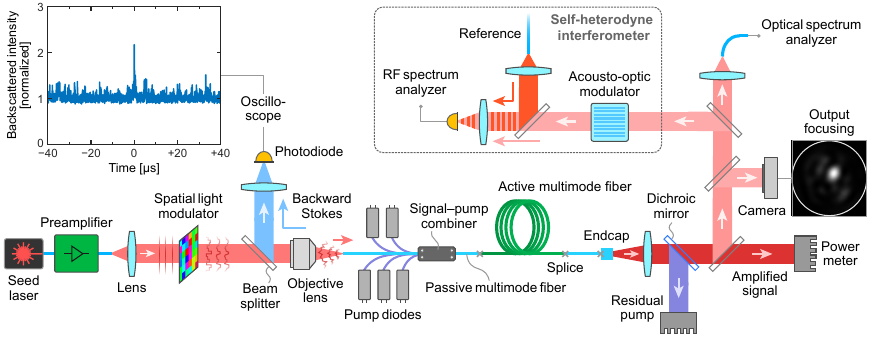}
		\caption{Schematic of our multimode fiber (MMF) amplifier with input wavefront shaping to tailor output beam profile. A single-frequency seed laser at wavelength $\lambda_{\mathrm S}=1064$~nm is preamplified to 10~W. The CW beam is expanded and wavefront shaped by a spatial light modulator (SLM), before coupling to a Yb-doped MMF amplifier pumped by five laser diodes. Amplified signal and residual pump are separated by a dichroic mirror, and their powers are measured by power meters. A camera records the near-field intensity distribution of amplified signal outside the MMF facet. The optical spectrum and linewidth of output signal are measured by an optical spectrum analyzer and a heterodyne interferometer. The SLM shapes the input wavefront to focus the output beam to a diffraction-limited spot. Right inset shows the focal spot at output power of 503~W. Left inset is the time trace of total backscattered intensity (normalized to its Rayleigh-scattering component), exhibiting a large spike due to SBS.}
		\label{fig:principle}
	\end{figure}

	\section{Results}    
	
	Figure~\ref{fig:principle} is a schematic of our master oscillator power amplifier (MOPA) configuration (see Methods for details). The linearly-polarized CW output of a single-frequency laser oscillator at wavelength $\lambda_{\mathrm S} = 1064$~nm is preamplified by two-stage Yb:doped SMF preamplifiers. Using a liquid-crystal SLM, we apply phase-only modulation to the linearly-polarized signal before coupling it to a Yb:doped MMF amplifier that supports 76 spatial modes. The doped fiber is cladding-pumped by five diode lasers in the wavelength range of 966--971~nm to amplify the signal via stimulated emission of Yb ions. The output signal is separated from the residual pump, and its power, optical spectrum and near-field intensity distribution are measured. We also monitor the backscattered light at the amplifier input end using a fast photodiode. The right inset of Figure~\ref{fig:principle} an optical image of the 503~W output beam from the amplifier that is focused to a chosen location beyond the fiber facet by shaping the input wavefront.
	
	\subsection{Stimulated Brillouin scattering}
	
	In the fiber amplifier, the signal being amplified acts as a pump for SBS, generating a backward propagating Brillouin Stokes light by emitting forward-propagating acoustic waves. 
    The Stoke light is amplified by both SBS and stimulated emission of excited Yb-ions as it propagates backward through the fiber. With increasing pump power, the time trace of the backscattered light intensity shows random spikes of duration $\sim$10--100 ns (left inset of Figure~\ref{fig:principle}). The sharp intensity fluctuations are attributed to SBS-induced dynamic instability~\cite{harrison1990evidence, gaeta1991stochastic, Agrawal2006,  panbiharwala2018experimental}, and represents a precursor for the onset of SBS~\cite{panbhiharwala2018investigation}. 
    Here we define the SBS instability threshold as the output signal power at which the maximum height of Stokes spikes is 1.5 times of the continuous background from Rayleigh scattering in backward intensity trace. This level is well below the conventional SBS threshold that is set as when the reflected power equals a few percent of the transmitted signal power~\cite{wisal2024theorySBS}. 
	
\begin{figure}
		\centering
		\includegraphics[width=\textwidth]{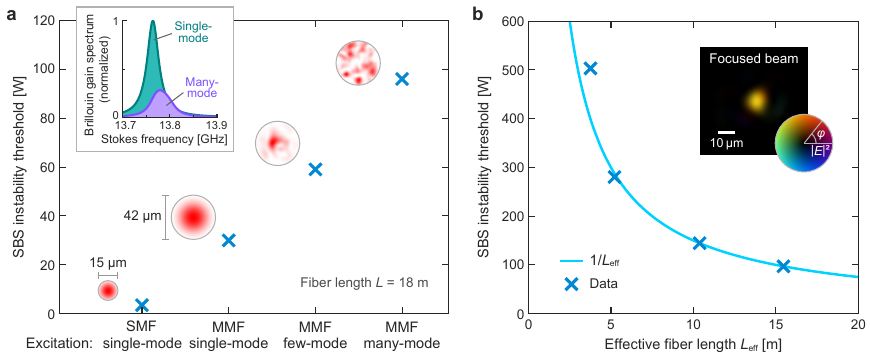}
		\caption{Stimulated Brillouin scattering (SBS) in fiber amplifiers. \textbf{a} Output signal power at the SBS instability threshold is estimated for a 15-\textmu m-core single-mode fiber (SMF) amplifier, and compared to for a 42-\textmu m-core multimode mode fiber (MMF) amplifier under single-mode (fundamental mode) excitation. Calculated intensity profile of the fundamental mode is shown as inset. SBS instability threshold for the MMF amplifier is measured with few-mode and multimode excitation, and output intensity distribution is shown next to the data points. Both SMF and MMF are 18~m long. Top left inset: normalized Brillouin gain spectrum calculated with equal excitation of all fiber modes (violet) is broader than fundamental-mode only excitation (green) in the MMF amplifier. \textbf{b} SBS instability threshold as a function of effective fiber length $L_{\mathrm{eff}}$ for the 42-\textmu m-core MMF with multimode excitation. Blue crosses are experimental data, and blue line shows the $1/L_{\mathrm eff}$ scaling. All data points are taken with output beam focused to a diffraction-limited spot at an axial distance of 50 \textmu m from the MMF end facet. Inset shows the spatial distribution of field intensity and phase across the focal spot at output signal power of 173~W. 76\% of total output power is concentrated within the focal area. Color represents the measured phase ($\phi$) and brightness the intensity ($|E|^2$).}
		\label{fig:sbsThreshResults}
	\end{figure}

	Compared to the standard SMF amplifier, the large cross section of our MMF amplifier lowers the signal intensity within the fiber core, thus reducing SBS. Similar to the conventional SBS threshold, the SBS instability threshold is expected to scale quadratically with the fiber core diameter (see supplementary material (SM) Sec.~2). If only the fundamental mode~(FM) is excited in our MMF with a 42-\textmu m core, the SBS threshold is approximately 8 times of that in a large-mode-area~(LMA) SMF amplifier with a 15-\textmu m core. From the previously measured SBS instability threshold in the SMF amplifier~\cite{panbiharwala2018experimental}, we estimate the SBS instability threshold for FM-only excitation in our MMF ampifier to be 24~W, as shown in Fig.~\ref{fig:sbsThreshResults}a. 
	
	Experimentally, it is difficult to realize FM-only excitation in the MMF amplifier, due to inevitable mode coupling in the fiber. When we use a lens of NA~$\approx$~0.03 to launch the signal into the MMF, the output beam is distorted from a smooth, symmetric profile (inset of Fig.~\ref{fig:sbsThreshResults}a), indicating both FM and a few high-order modes~(HOMs) are excited. The measured SBS instability threshold for the few-mode excitation is 64~W. 	
	To excite more HOMs in the MMF amplifier, we replace the lens with a microscope objective of NA~=~0.13, slightly larger than the fiber NA~=~0.1. With the input signal tightly focused onto the front facet of the MMF, the output beam is speckled, as seen in the inset of Fig.~\ref{fig:sbsThreshResults}a. The measured SBS instability threshold for multimode excitation is 97~W.

	\subsection{Multimode SBS theory}
	
	The four-fold increase of SBS instability threshold (24 W to 97 W) by multimode excitation within the same MMF can be explained by our semi-analytic theory of SBS in the MMF amplifier~\cite{chen2023mitigating, wisal2024theorySBS, wisal2024optimal}. When the input light is distributed over multiple fiber modes, the forward-propagating signal in each mode will be Brillouin-scattered to backward-propagating Stokes in all modes. This process is mediated by acoustic modes in the fiber. The Stokes is frequency downshifted from the signal by the acoustic frequency $\Omega$, leading to a peak in the Brillouin scattering spectrum. The scattering coefficient for signal in the $l$-th mode and Stokes in the $m$-th mode,  $g_{\mathrm B}^{(m,l)}(\Omega)$, depends on the spatial overlap of the signal and Stokes mode profiles with various acoustic modes in the fiber. Generically, intermodal scattering ($m\neq l$) is significantly weaker than the intramodal scattering ($m=l$) due to smaller acousto-optic overlap. Each mode pair of $l$ and $m$ typically has the largest spatial overlap with a single acoustic mode, making $g_{\mathrm B}^{(m,l)}$ peaked at this acoustic mode frequency $\Omega$. Overall Brillouin scattering to the $m$-th mode from signals in all fiber modes is: 
	\begin{equation}
	G_{\mathrm B}^{(m)}(z)= \sum_l g_{\mathrm B}^{(m,l)}(\Omega) \, {P}_{\mathrm S}^{(l)}(z),
	\end{equation}
	where ${P}_{\mathrm S}^{(l)}$ is the signal power in the $l$-th mode, and $z$ is the longitudinal coordinate along the fiber axis. While propagating through the MMF amplifier, the signal power grows via stimulated emission, enhancing the Brillouin scattering. The growth rate of Stokes power in the $m$-th mode $P^{(m)}_{\mathrm B}$ is determined by $G_{\mathrm B}^{(m)}$,
	\begin{equation}
	\frac{d P^{(m)}_{\mathrm B}}{-dz} = G_{\mathrm B}^{(m)} \, P^{(m)}_{\mathrm B}(z) \, .   
	\end{equation} Due to the exponential nature of growth, the total Stokes power over all modes is dominated by the mode that has the maximum value of $G_{\mathrm B}^{(m)}$.
	
	$G_{\mathrm B}^{(m)}$ depends on the signal power distribution among the fiber modes. We consider three cases below. First, all the signal power is in the FM, $l=1$. $g_{\mathrm B}^{(1,1)}$ is larger than $g_{\mathrm B}^{(m,1)}$ for any $m > 1$, due to better spatial overlap between optical and acoustic modes. Thus Brillouin scattering is strongest for the Stokes in the FM ($m=1$), mediated by the lowest-order acoustic mode in the fiber. Next, a single higher-order mode (HOM), $l > 1$, is excited. As mentioned earlier, intramodal Brillouin scattering ($m=l$) is stronger than intermodal scattering ($m \neq 1$), $g_{\mathrm B}^{(l,l)} > g_{\mathrm B}^{(m,l)}$ for any $m \neq l$. However, smaller acousto-optic overlap makes $g_{\mathrm B}^{(l,l)} < g_{\mathrm B}^{(1,1)}$, and the SBS threshold is somewhat higher than that with FM-only excitation. Finally, if input power is distributed among all fiber modes, $G_{\mathrm B}^{(m)}$ is a sum of intramodal and intermodal scattering coefficients weighted by reduced signal power ${P}_{\mathrm S}^{(l)}$ in individual modes. Because intermodal scattering is weaker than intramodal and $g_{\mathrm B}^{(m,l)}$ for different mode pairs are peaked at varying frequency $\Omega$, $G_{\mathrm B}^{(m)}$ is spectrally broadened and has notably lower peak value than single-mode excitation (inset of Fig.~\ref{fig:sbsThreshResults}a). As the peak value dictates the exponential growth of Stokes power, the SBS threshold is significantly higher for multimode excitation than for the excitation of any single mode.

	\subsection{Pump depletion and gain saturation}
	
	In the high-power MMF amplifier like ours, gain saturation and pump depletion are strong. These effects determine the signal power distribution throughout the fiber, which has a significant impact on the Stokes growth rate. While our earlier modeling elucidated the advantages of multimode excitation for reducing SBS in passive and active fibers~\cite{wisal2024theorySBS}, it did not take into account these effects. To compute $G_{\mathrm B}^{(m)}$ including these effects, we extend that theory as follows. We first calculate the signal power in each mode ${P}_{\mathrm S}^{(l)}(z)$ throughout the fiber, taking into account gain saturation and pump depletion in the MMF amplifier~\cite{wisal2024theoryTMI}. At or below the SBS instability threshold, the Stokes power is much lower than the signal power, thus feedback due to SBS is neglected when computing the signal power and pump power distribution in the MMF amplifier (see SM Sec.~2 for more detail). 	
	
	Numerical simulation of our MMF amplifier using the improved model (with parameters listed in SM table 1) reveals that the pump power (at wavelength $\lambda_{\mathrm P} = 971$ nm) is mostly absorbed ($\sim 90 \%$) in the first 6~m of the Yb:doped fiber. Beyond this distance, the signal power does not grow significantly (Fig.~S2). In Fig.~\ref{fig:sbsThreshResults}a, the length of Yb:doped MMF plus a short undoped MMF spliced to its distal facet is $L$~=~18~m, and in the last two thirds of the fiber the signal power is nearly constant, but the Stokes power grows rapidly because of high signal power. Hence, we cut back the MMF to mitigate SBS. From the physical fiber length $L$ and total signal power $P_{\mathrm S}(z) = \sum_l {P}_{\mathrm S}^{(l)}(z)$, we compute the effective fiber length for SBS: 
	$L_{\mathrm{eff}}={\int^L_0{P_{\mathrm S}(z) \, dz}}/{P_{\mathrm{S}}(L)}$.
	
	Figure~\ref{fig:sbsThreshResults}b shows that the measured SBS instability threshold increases rapidly as $L_{\mathrm{eff}}$ decreases (blue crosses). The blue curve shows that SBS threshold scales inversely with $L_{\mathrm{eff}}$, as expected from our theory (SM Sec.~2). 
	At the shortest length $L_{\mathrm{eff}}$ = 3.7~m, the SBS instability threshold reaches 503~W experimentally. 
	
	\subsection{Output beam shaping}
	
	All data points in Fig.~\ref{fig:sbsThreshResults}b are taken with the output beam focused to a chosen spot outside the distal facet of the MMF by optimizing the input phase front with the SLM (see Methods). The inset is an optical image of focal spot taken at 173~W output signal power. With phase-only modulation of input signal,  76\% of total output power is concentrated within the focal area. The near-field focusing of output beam is achieved through interference of signal in many fiber modes, and the SBS instability threshold is enhanced by multimode excitation in the amplifier. 
	
	The ability of controlling output beam profile by input wavefront shaping relies on that the signal bandwidth is narrower than the spectral correlation width of MMF amplifier. The latter characterizes how fast the output field pattern decorrelates with frequency detuning for a fixed input wavefront. Despite gain saturation and pump depletion in our MMF amplifier, its spectral correlation width barely changes from that of the passive MMF~\cite{rothe2025output}. With increasing fiber length, the spectral correlation width decreases; but even for the longest MMF of our amplifier, the correlation width exceeds 1~GHz, which is well above the signal bandwidth (see next section). The amplified signal is spatially coherent; that is, the relative phase of output fields between any two positions is time-invariant. Consequently, the speckle pattern created by multimode interference at any frequency within the signal bandwidth is almost identical, leading to high intensity contrast as seen in the inset of Fig.~\ref{fig:sbsThreshResults}a. 
	
	The spatial coherence of amplified signal allows us to generate the same output pattern, e.g., focusing to a chosen location for all frequency components, by imposing a single wavefront on the input beam. To measure the phase of output field, we perform an interferometric experiment (see Methods). The inset of Fig.~\ref{fig:sbsThreshResults}b shows the measured phase is constant across the focal spot, confirming diffraction-limited focusing.  
	
	\begin{figure}[h]
		\centering
		\includegraphics[width=\textwidth]{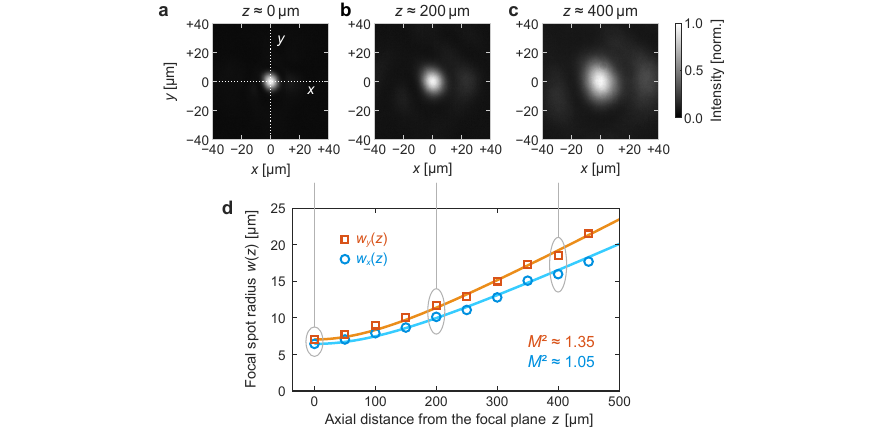}
		\caption{Beam propagation factor of focal spot at MMF amplifier output. \textbf{a} Intensity profile of a diffraction-limited spot created at 300~\textmu m from the MMF distal facet by shaping phase front of input signal. $76\%$ of total output power is concentrated inside the focal area. \textbf{b},\textbf{c} Output beam profiles recorded at axial distance $z \approx$ 200~\textmu m (\textbf{b}) and 400~\textmu m (\textbf{c}) from the focal plane, respectively. \textbf{d} Focal spot radii along $x$ and $y$ axes (perpendicular to fiber axis $z$), $w_x$ and $w_y$, increase with axial distance $z$. Squares and circles mark experimental data, solid curves represent the fit with Eq.~\ref{eq:M2} that gives $M_x^2 \approx 1.05$ and $M_y^2 \approx 1.35$.}
		\label{fig:beamShaping}
	\end{figure}
	
	To track axial evolution of the focal spot, we image the transverse intensity distribution at different axial planes. Three exemplary images are presented in Fig.~\ref{fig:beamShaping}a,b,c, revealing the spot radius increases with axial distance $z$ from the focal plane. We extract the intensity profiles in $x$ and $y$ axes, and fit with Gaussian envelop to obtain the spot radius $w_x$ and $w_y$ (see Methods). Figure~\ref{fig:beamShaping}d shows that $w_x$ and $w_y$ grow with $z$. We fit the $z$ dependence with 
	\begin{equation}
	w(z)= w(0) \sqrt{1+ \left( \frac{M^2 \, \lambda_{\mathrm S} \, z}{\pi \, n \,  [w(0)]^2 } \right)^2} \,
	\label{eq:M2} 
	\end{equation} 
	where $w(0)$ is the focused spot radius at the focal plane, $\lambda_{\mathrm S}$ is the signal wavelength in vacuum, and $n\approx 1.45$ is the refractive index of the endcap that the output beam propagates. Eq.~\ref{eq:M2} fits the data well and give the beam propagation factors $M_x^2 \approx 1.05$ and $M_y^2 \approx 1.35$.

	\subsection{Efficiency and linewidth}
	
\begin{figure}
		\centering
		\includegraphics[width=\textwidth]{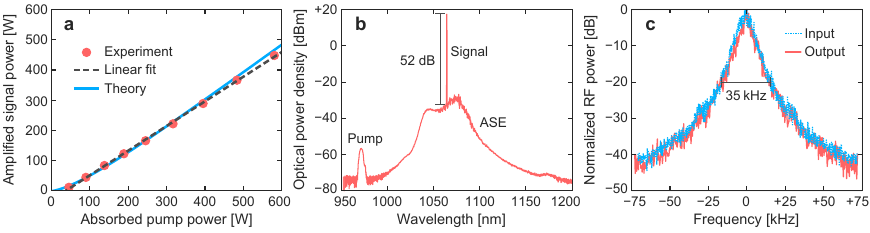}
		\caption{High-power MMF amplifier with narrow linewidth. The fiber core diameter is 42~\textmu m and effective length $L_{\mathrm {eff}}$ = 3.7~m. \textbf{a} Amplified signal power as a function of absorbed pump power gives the slope efficiency of $82$\%. Red circles are experimental data, black dashed line is a linear fit of data, and blue curve from our theoretical model. \textbf{b} Optical spectrum of amplifier output at 453~W power shows the amplified signal peak at 1064~nm on top of a broad amplified ASE band. The signal-to-ASE peak ratio is 52~dB. The small peak centered at 971~nm is from residual pump. \textbf{c} Heterodyne spectra for input (blue) and output (red) signals of the MMF amplifier at the output signal power of 503~W. Full width at -20~dB of maximum is 35~kHz for both input and output signals, indicating no detectable spectral broadening by the MMF amplifier. After subtracting the reference linewidth, the 20-dB width of input/output signal is 18~kHz.  }
		\label{fig:efficiencySpectrum}
	\end{figure}

	We further characterize the performance of high-power MMF amplifier with the shortest $L_{\mathrm{eff}}$ = 3.7~m. First, we measure the output signal power and residual pump power. The power of amplified signal is given by the difference between input and output signal power, $P_{\mathrm S}^{\mathrm (amp)} = P_{\mathrm S}^{\mathrm (out)}-P_{\mathrm S}^{\mathrm (in)}$. The absorbed pump power is obtained by subtracting the residual pump power at the amplifier output from the pump power launched into the MMF, $P_{\mathrm P}^{\mathrm (abs)} = P_{\mathrm P}^{\mathrm (in)}-P_{\mathrm P}^{\mathrm (out)}$. Figure~\ref{fig:efficiencySpectrum}a shows $P_{\mathrm S}^{\mathrm (amp)}$ increases almost linearly with $P_{\mathrm P}^{\mathrm (abs)}$. A linear fitting gives the slope efficiency of $82\%$. The experimental data agree with theoretical prediction of our model. 
	
	We also measure the output spectrum of the MMF amplifier with an optical spectrum analyzer. In Fig.~\ref{fig:efficiencySpectrum}b, the signal appears as a narrow peak at $\lambda_{\mathrm S}$ = 1064~nm. It sits on top of a broad amplified spontaneous emission~(ASE) band. The peak ratio of amplified signal to ASE is 52~dB. 
	
	The signal linewidth, which is too narrow to be resolved by the optical spectrum analyzer, is measured by a heterodyne interferometer (see Methods). 
	Figure~\ref{fig:efficiencySpectrum}c shows the heterodyne peak resulting from beating of the amplified signal and a reference (red curve). The full width at -20dB of the maximum is $\Delta \nu_{\mathrm M}=35$~kHz. Subtracting the reference width of $\Delta \nu_{\mathrm R} = 17$~kHz (see Fig.~\ref{fig:seedLinewidth}), the 20-dB width of the output signal is $\Delta \nu_{\mathrm S}= \Delta \nu_{\mathrm M} - \Delta \nu_{\mathrm R} = 18$~kHz. 
	It corresponds to a Lorentzian line of 3~dB bandwidth (full width at half maximum) $\sim1$~kHz.

	For comparison, we measure the linewidth of input signal to the MMF amplifier using the same method. As shown by the blue curve in Figure~\ref{fig:efficiencySpectrum}c, the input signal has nearly identical width as the output (red curve), thus coherent multimode amplification does not cause detectable spectral broadening of the signal. As mentioned earlier, a common approach to mitigating SBS in SMF amplifiers is by broadening the input signal linewidth to tens of GHz. In our scheme, the 3~dB signal linewidth is 6 orders of magnitude narrower, thus the temporal coherence length is 6 orders of magnitude longer in the MMF amplifier.

	\section{Discussion and conclusion}

	Input wavefront shaping allows to simultaneously mitigate SBS in the MMF amplifier and control the output beam profile. In the current experiment, phase-only modulation of input wavefront limits output focusing efficiency, about $20-30\%$ of power is outside the focal spot. Complete focusing can be achieved with both amplitude and phase modulation of the input signal~\cite{rothe2025output}. Full-field modulation will also generate values of $M^2$ closer to unity. Experimentally this can be realized with two separate phase masks, or with double bounces from different regions of a phase-only SLM. Phase modulations, in principle, do not introduce power loss. 
	
	In addition to output focusing, input wavefront shaping can generate different output profile of the MMF amplifier, which will be useful for laser welding and material processing. Even in the presence of strong nonlinearity, gain saturation and pump depletion, as long as the amplifier operates below the instability threshold, optimal input wavefront can be found by minimizing the difference between the measured output beam shape and the target~\cite{chen2024exploiting}. In this work, the spatial profile of a single linear polarization of the output beam is controlled by input wavefront shaping. Although the input signal is linearly polarized, polarization mixing in the MMF causes depolarization. However, it is possible to control output beam profiles for two orthogonal polarizations by shaping both input polarizations. This can be done by separating two orthogonal polarizations of the input signal and shaping their field patterns separately before combining them and coupling to the MMF~\cite{rothe2025output}.  
	
	Further power scaling of our MMF amplifier is possible by shortening the fiber length to increase the SBS threshold. However, the pump light would not be fully absorbed, lowering the power efficiency of the MMF amplifier. A more efficient way of increasing the SBS threshold is enlarging fiber core size. Even if only the fundamental mode is excited, the SBS threshold scales quadratically with the core diameter (SM Sec.~2). For multimode excitation, Brillouin spectrum broadening is more pronounced, since there are more modes in a MMF with larger core, leading to higher enhancement of the SBS threshold. While our scheme does not rely on specialty fibers, it can be applied to MMFs with lower optical nonlinearity and microstructured fibers with large cross-section for higher-power operation.     
	
	More generally, our scheme can be extended to mitigating other detrimental nonlinear effects in high-power fiber amplifiers, e.g., transverse mode instability, stimulated Raman scattering, modulation instability. Our scheme of spatial and modal spread of the signal may be combined with temporal stretch of optical pulses for high-peak-power amplification. However, if the signal bandwidth exceeds the spectral correlation width of the MMF amplifier, input spatial shaping needs to be replaced by spatiospectral shaping to control the spatiotemporal profile of the output pulse~\cite{mounaix2020time, cruz2022synthesis, yessenov2022vector}.

	\section{Methods}
	
	\subsection{Multimode fiber amplifier}
	
	The oscillator is a single-frequency fiber laser (NP Photonics RFLM-100-1-1064 Rock module) that produces a CW linearly-polarized beam of power up to 100~mW at wavelength $\lambda_{\mathrm S}$ = 1064~nm. The oscillator output is split into two: one serves as the seed for fiber amplifiers, the other as the reference for interferometic measurements. The seed of power 25~mW  is launched into a two-stage preamplifier with Yb-doped SMFs and amplified to 10~W. The output beam is then wavefront-shaped by a phase-only liquid-crystal SLM (Santec SLM-300), which is placed at the focal plane of the input facet of a double-clad step-index MMF without Yb doping (Coherent FUD-4693). This fiber, labeled signal fiber, has a core diameter 42~$\mu$m and numerical aperture~(NA) 0.1. It supports 76 guided modes at the signal wavelength $\lambda_{\mathrm S} = 1064$~nm. The proximal facet of the signal fiber is angle-cleaved ($\sim 8^{o}$) to avoid spurious reflection. Its distal end is combined with the SMF pigtails of five laser diodes (model number ?), each providing up to 125~W power. With increasing electric current, the pump wavelength $\lambda_{\mathrm P}$ gradually shifts from 966 nm to 971 nm. The output MMF of signal-pump combiner is spliced to a Yb-doped MMF (Coherent FUD-4715) with same parameters as the signal fiber (core, first cladding, and second cladding diameters are 42, 570, and 650~\textmu m, respectively). The distal facet of the doped MMF is spliced to a short undoped MMF that is terminated by an antireflection-coated endcap. An optical isolator is placed between the seed laser and the preamplifier, a second one between the preamplifier and the SLM, and a third one between the SLM and the MMF amplifier.  

    
	\subsection{Wavefront shaping}
	
	The SLM area covering the NA of MMF core is divided into 256 macropixels, each consisting of $72 \times 72$ SLM pixels of area $9 \times 9$~\textmu m$^2$. We adjust the phase of each macropixel sequentially to maximize the output intensity at the chosen location. We start from the center macropixel, scan its phase $\phi$ from 0 to $2\pi$, and measure the power within the focal area of diameter given by twice the full width at half maximum of the maximum intensity.
    Since the focused power typically varies sinusoidally with the phase, the measured power is fitted by an objective function ${\mathrm cos}(\phi - \phi_0) + \text{const.}$ with $\phi_0$ representing the phase for maximum power. After setting the phase of that macropixel to $\phi_0$, we repeat the phase-optimization procedure for neighboring macropixels, proceeding in a spiral pattern outward from the center, until the optimal phases for all macropixels are determined. 
    After the optimization, we characterize the focusing efficiency by the ratio of power within the focal area and the total power. 
    To determine the radius $w_{x,y}$ of focal spot, we fit the intensity distribution along $x$ or $y$ axis with a Gaussian envelop and set $w_{x,y}$ where the intensity drops to $e^{-2} $ of the maximum at the center.    
	
	\subsection{Optical characterizations}
	
	The output signal is divided into multiple beams for spatial, spectral and temporal characterization. The signal power is measured by a water-cooled power meter (Ophir Spiricon L1500W-BB-50). The optical spectrum is recorded by an optical spectrum analyzer (Yokogawa AQ637OD). The temporal variation of signal intensity is detected by a fast photodiode (Thorlabs PDA20CS) that is connected to a high-speed oscilloscope (Keysight DSOX3014T). Spatial intensity distribution of the output signal on a transverse plane at a distance of 50~\textmu m beyond the fiber and endcap interface is imaged by a lens to a CCD camera (Allied Vision Manta 6419B-NIR). A linear polarizer (Thorlabs LPIREA 100-C) is placed before the camera to select one polarization. To measure its phase, the linear-polarized signal interferes with a reference beam from the seed laser. The spatial phase distribution of the signal field is extracted from the off-axis hologram.    
	
	The spectral linewidth of the seed laser is measured by a delayed self-heterodyne interferometer. Schematics are shown in Fig.~\ref{fig:principle} and Fig.~\ref{fig:expSchematic}. The output beam is split into two. One of them is modulated by an acousto-optic modulator~(AOM) at frequency 110~MHz. The other beam is delayed by propagating through a 25-km SMF. The two incoherent beams are then recombined by a beam splitter and the total intensity is measured by a fast photodiode which is connected to an electrical spectrum analyzer (Keysight N9030B-PXA). The 20-dB width of the heterodyne peak at 110~MHz is 34~kHz (SM Fig.~\ref{fig:seedLinewidth}). Since it is equal to two times the laser linewidth, thus the 20-dB linewidth of seed laser is 17~kHz.   
	
	To measure the MMF amplifier linewidth, a fraction of the output signal is modulated by the AOM at frequency 110~MHz. It is then combined with a reference beam from the seed laser. To remove its coherence with the signal, the reference is delayed via a 25-km SMF. The sum of modulated signal and delayed reference is measured by a fast photodiode and an electrical spectrum analyzer. The width of the heterodyne peak at 110~MHz is equal to the sum of the signal linewidth and reference linewidth. 	
	

	\subsection*{Acknowledgements}
	We thank Filipe~M.~Ferreira, Nicholas~Bender, Owen~D.~Miller, Ori~Henderson-Sapir, Stephen~C.~Warren-Smith, Darcy~L.~Smith, Linh~V.~Nguyen, David~J.~Ottaway, Heike~Ebendorff-Heidepriem, Michel~J.~F.~Digonnet, and John~Ballato for fruitful discussions. We acknowledge the technical assistance from Nicholas~Bernardo and Vincent~Bernardo at Yale University and the support from AFL, Thorlabs, Lightel, and NP Photonics. This work is supported by the Air Force Office of Scientific Research (AFOSR) under Grant FA9550-24-1-0129 (H.C. and A.D.S.) and the German Research Foundation (DFG) under Grant RO~7348/1-1~(S.R.).  \\
	
    \subsection*{Author contributions}
    H.C. proposed the idea and initiated this project; S.R., C.-W.C., and P.A. designed and built the multimode fiber amplifier system, performed the experiments in collaboration with M.E., K.L., and N.V, and analyzed the data in collaboration with K.W., under the supervision of H.C.; K.W. developed the theory and numerical simulations under the supervision of A.D.S.; S.R., C.-W.C., K.W., A.D.S., and H.C. wrote the manuscript with input from all authors.
    
	\subsection*{Competing interests}
	The authors declare no competing interests.
	
	\subsection*{Additional information}
	\textbf{Supplementary information}
	

	\clearpage
	\renewcommand{\appendixname}{Supplementary Material}
	\renewcommand{\appendixpagename}{Supplementary Material}
	\renewcommand{\appendixtocname}{Supplementary Material}
	\begin{appendices}
		\renewcommand{\thesection}{\arabic{section}}
		\setcounter{figure}{0}
		\renewcommand{\thefigure}{S\arabic{figure}}
		\setcounter{table}{0}
		\renewcommand{\thetable}
		{S\arabic{table}}

\section{Experimental setup and spectral linewidth of the seed laser}

\begin{figure}[h]
    \centering
    \includegraphics[width=\textwidth]{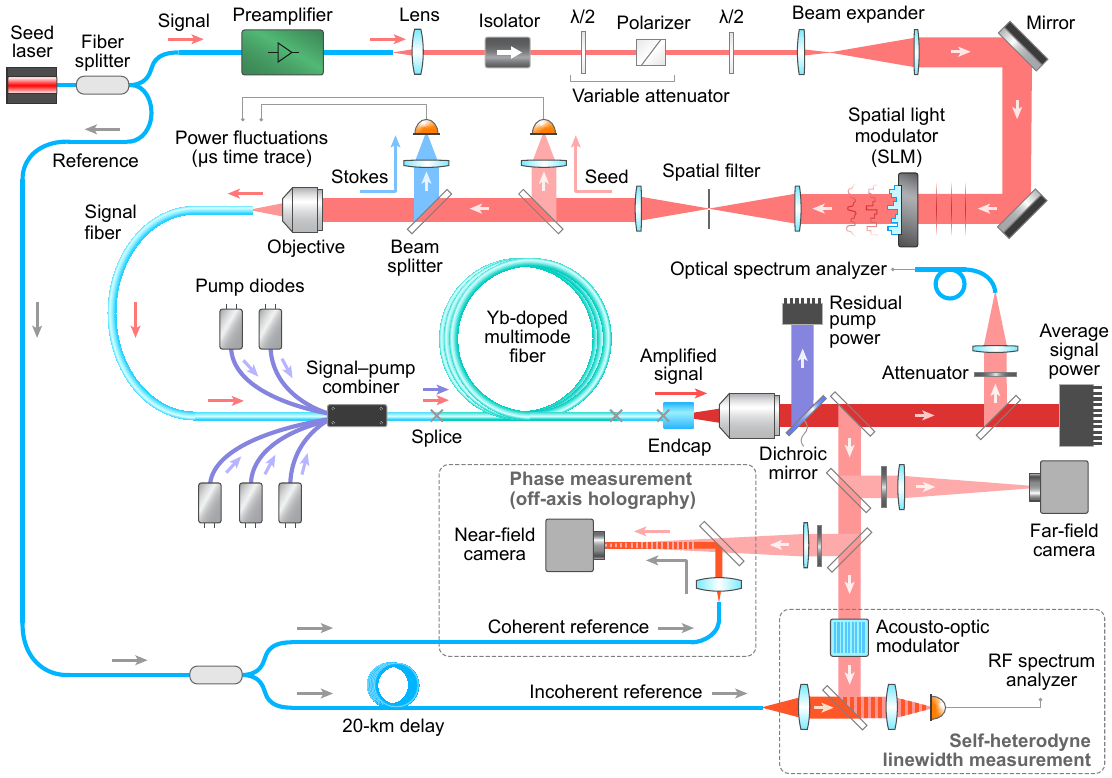}
    \caption{Detailed schematic of our MMF amplifier and characterization setup.}
    \label{fig:expSchematic}
\end{figure}

\begin{figure}[h]
    \centering
    \includegraphics[width=\textwidth]{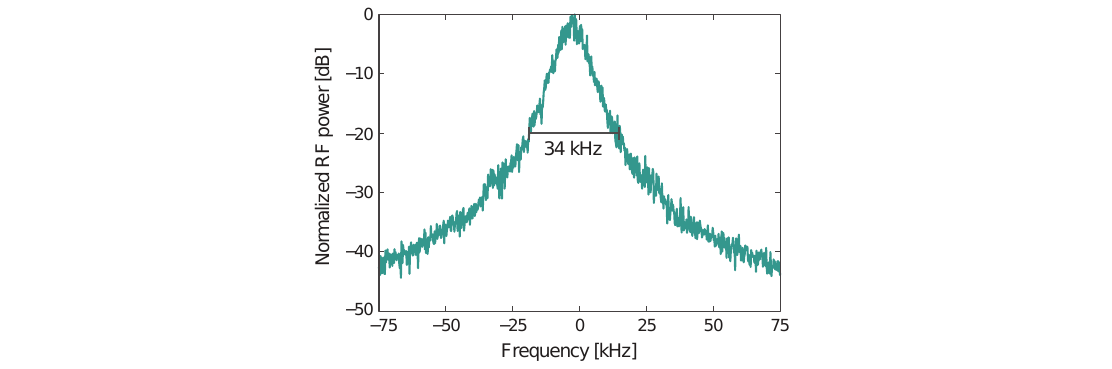}
    \caption{Heterodyne spectrum for measuring the spectral linewidth of the seed laser. The full width at $-$20~dB of maximum is 34~kHz.}
    \label{fig:seedLinewidth}
\end{figure}

\section{Theoretical model}

We have derived a semi-analytic theory of vector optical modes coupled via scalar acoustic waves to describe SBS in highly multimode fiber (MMF) amplifiers, extending previous studies on passive multimode fibers and single-mode-fiber (SMF) amplifiers~\cite{smith2011mode, chen2023mitigating, wisal2024theorySBS, wisal2024optimal}. Here we fully take into account optical gain saturation and pump depletion, which were ignored in our previous studies~\cite{wisal2024theorySBS,chen2023mitigating}. We consider only Stokes amplification by SBS, and not stimulated emission of Yb ions, because the Brillouin gain is orders of magnitude higher than stimulated emission below the SBS threshold.  Our theoretical model elucidates the underlying mechanism of the SBS suppression and can account for the observed scaling of SBS threshold in the MMF amplifier. 

As mentioned in the main text, coherent multimode excitation in an MMF amplifier offers two key advantages over the SMF counterpart. First, the larger fiber core reduces light confinement, and weaker signal intensity within the core reduces optical nonlinearity. Quantitatively, the SBS threshold scales linearly with the fiber core area. Second, the SBS threshold is further increased by highly multimode excitation in the MMF amplifier~\cite{chen2023mitigating, wisal2024theorySBS}. The Brillouin gain is reduced, both because intermodal scattering is weaker than intramodal scattering, and effective spectrum broadens due to differential frequency displacement of Brillouin scattering peaks for different mode pairs. Hence SBS threshold is higher when the signal excites more modes in the fiber. These effects were predicted and observed in passive fibers \cite{chen2023mitigating,wisal2024theorySBS} previously; our improved theory, which includes gain saturation and pump depletion, shows that the physical origin of the reduction of Brillouin gain is the same in MM amplifiers.  This allows us to make a solid scaling prediction for the variation of the threshold with the effective length of the fiber.
  
The SBS threshold can be further increased by optimizing the fiber length. Since the Stokes power grows exponentially in the backward propagation through the fiber amplifier, shortening the fiber length will reduce SBS. However, the output signal power may be lower if the signal amplification is truncated. 
The optimal length of a fiber amplifier is given by the minimum length for nearly complete absorption of the pump light. We numerically find such length by simulating the coupled evolution of pump power and signal power in a MMF amplifier~\cite{wisal2024theoryTMI}, taking into account pump depletion and gain saturation. 

Consider a co-pumped Yb-doped MMF amplifier. The fiber core is doped with Yb ions that are pumped to amplify the signal via stimulated emission. Below the SBS instability threshold, the Stokes power is much lower than the signal power, allowing us to ignore SBS when computing  signal power and pump power throughout the MMF amplifier. The field amplitude $A_{\mathrm S}^{(l)}$ of signal in the $l$-th fiber mode at position $(\vec{r}_\perp,z)$ satisfies the following equation:

\begin{equation}
\frac{dA_{\mathrm S}^{(m)}(z)}{dz} = \sum_l e^{i(\beta_l-\beta_m)z} \, g_{\mathrm E}^{(m,l)}(z) \, A_{\mathrm S}^{(l)} (z)\:\:,\:\:   g_{\mathrm E}^{(m,l)}(z) =  \frac{g_0}{2}\int d \vec{r}_\perp \frac{\psi^*_m (\vec{r}_\perp) \, \psi_l (\vec{r}_\perp)}{1+I_{\mathrm S}(\vec{r}_\perp,z)/I_{\mathrm sat}(z)}. 
\label{Eq:signalamp}
\end{equation}
Here, $\beta_m$ ($\beta_l$) is the propagation constant of mode $m$ ($l$), $g_{\mathrm E}^{(m,l)}$ is the gain coefficient for mode $m$ from mode $l$, $g_0$ is unsaturated gain coefficient which is assumed to be mode independent, $\psi_l$ ($\psi_n$) is the transverse field profile of fiber mode $l$ ($m$), $I_{\mathrm S}$ is the signal intensity given by $I_{\mathrm S}(\vec{r}_\perp,z)={|\sum_l A_l(z) \, \psi_l(\vec{r}_\perp) \, e^{i\beta_l z}|}^2$, and $I_{\mathrm sat}$ is the gain saturation intensity. $g_{\mathrm E}^{(m,l)}$ depends on the spatial overlap of $\psi_m$ and $\psi_l$ weighted by the intensity-dependent gain saturation over Yb doped area of the fiber. The cross modal gain $m \neq l$ is non-zero because of spatial hole burning, i.e., transverse-varying gain saturation by $I_{\mathrm S}(\vec{r}_\perp,z)$. As a result, the equations for different modal amplitudes are coupled, unlike the MMF amplifier with unsaturated or weakly saturated gain. Both $I_{\mathrm sat}$ and $g_0$ depend on the pump power $P_{\mathrm P}$, which evolves along the fiber as:
\begin{equation}
\frac{dP_{\mathrm P}(z)}{dz} = -g_{\mathrm P}(z) P_{\mathrm P}(z)\:\:,\:\:    g_{\mathrm P}(z) = \frac{\rho_{\mathrm t}}{A_{\mathrm c}} \int d\vec{r}_\perp \, \left[ \sigma_{\mathrm P}^{\mathrm (a)} \rho_{\mathrm l}(\vec{r}_\perp,z) - \sigma_{\mathrm P}^{\mathrm (e)} \rho_{\mathrm u}(\vec{r}_\perp,z)\right].
\label{Eq:pumpdecay}
\end{equation}
Here $g_{\mathrm P}(z)$ is the absorption coefficient of pump light, $\rho_{\mathrm t}$ is the density of Yb ions,  $\rho_{\mathrm u}$ ($\rho_{\mathrm l}$) is the fraction of Yb ions in the upper (lower) energy level, $\sigma_{\mathrm P}^{(a)}$ ($\sigma_{\mathrm P}^{(e)}$) is the absorption (emission) cross-section of Yb ions at the pump wavelength $\lambda_{\mathrm P}$. The integral is over Yb doped area of the fiber core, and $A_{\mathrm c}$ is the area of pump core (inner cladding of double-clad fiber). Note that we consider a single equation for the sum of pump power in all modes, because the pump light is incoherent and fills the pump core uniformly~\cite{paschotta1997ytterbium}. Finally, the fraction of upper ($\rho_{\mathrm u}$) and lower ($\rho_{\mathrm l}=1-\rho_{\mathrm u}$) level density at steady state is obtained from the rate equations of Yb ions~\cite{wisal2024theoryTMI}. We solve Eq.~\ref{Eq:signalamp} and Eq.~\ref{Eq:pumpdecay} together with the expression for $\rho_{\mathrm u}$ using a finite-difference method such as Euler or Runge--Kutta method.

\begin{figure}[t]
    \centering
    \includegraphics[width=\textwidth]{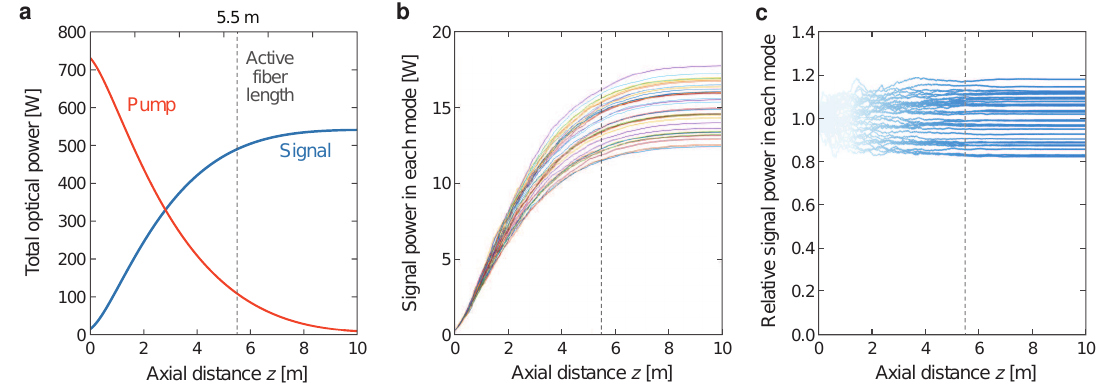}
    \caption{ (a) Numerically calculated pump and signal power evolution in the MMF amplifier with parameters given in Table~\ref{tab:B1}. Vertical dashed line marks shortest active fiber length of 5.5~m, where the signal power (blue solid curve) grows from 10~W to 490~W, and the pump power (orange dotted curve) reduces from 725~W to 98~W. (b) Signal power evolution in individual fiber modes shows weak mode-dependent gain. (c) Relative modal power, given by the ratio of signal power in each fiber mode and that averaged over all fiber modes, varies with $z$ in the first half of the active fiber and becomes nearly $z$ independent in the second half. The brightness of each curve is proportional to the total signal power.  highlighting the region where SBS is significant. }
    \label{fig:pvsz}
\end{figure}

Figure~\ref{fig:pvsz}a shows the pump power $P_{\mathrm P}(z)$ decays as the total signal power $P_{\mathrm S}(z) = \sum_l P_{\mathrm S}^{(l)} (z) $ grows along the MMF amplifier with parameters listed in Table ~\ref{tab:B1}.  After $\sim6$~m propagation in the fiber, the signal gain is saturated and nearly all the pump power is depleted. Beyond this point, the signal power grows very little and eventually decreases gradually, due to signal reabsorption by unexcited Yb ions. Thus the optimal length of our MMF amplifier is around 6 m at pump wavelength $\lambda_{\mathrm P} = 971$ nm. Experimentally, the shortest length of Yb-doped MMF is 5.5 meter, and the SBS instability threshold is 503 W. 

Figure~\ref{fig:pvsz}b shows the signal power $P_{\mathrm S}^{(l)}$ of individual mode grows at slightly different rate, as the gain saturation varies from mode to mode. The power fraction $\tilde{P}_{\mathrm S}^{(l)}(z) = P_{\mathrm S}^{(l)}(z) / P_{\mathrm S}(z)$ initially varies with $z$, then remains almost constant after $z > 3$ m, where the signal power is high and Brillouin Scattering is strong (Fig.~\ref{fig:pvsz}c). Eq.~(1) in the main text becomes
\begin{equation}
G_{\mathrm B}^{(m)}(z) = P_{\mathrm S}(z) \, \sum_l g_{\mathrm B}^{(m,l)} \, \tilde{P}_{\mathrm S}^{(l)} = P_{\mathrm S}(z) \, \tilde{G}_{\mathrm B}^{(m)} \,  
\end{equation}
where $\tilde{G}_{\mathrm B}^{(m)} \equiv \sum_l g_{\mathrm B}^{(m,l)} \, \tilde{P}_{\mathrm S}^{(l)} $ is approximately independent of $z$ for $z > 3$ m where Brillouin gain is high. Hence,
\begin{equation}
P_{\mathrm B}^{(m)}(z) \simeq P_{\mathrm B}^{(m)}(L) \, e^{\tilde{G}_{\mathrm B}^{(m)} \, \int_z^L P_{\mathrm S}(z) \, dz} \, , 
\end{equation}
where $P^{(m)}_{\mathrm B}(L)$ is the Stokes power in the $m$-th mode at the distal end of the fiber ($z=L$) from spontaneous Brillouin scattering. 
The Stokes power at the proximal end $(z=0)$, summed over all modes, is dominated by the mode with largest $G_{\mathrm B}^{(m)}$. It is approximated by 
\begin{equation}
P_{\mathrm B}^{(m)}(0) \simeq P_{\mathrm B}^{(m)}(L) \, e^{\tilde{G}_{\mathrm B}^{(m)} \, \int_0^L P_{\mathrm S}(z) \, dz} =  P_{\mathrm B}^{(m)}(L) \, e^{\tilde{G}_{\mathrm B}^{(m)} \, P_{\mathrm S}(L) \, L_{\mathrm{eff}}} \, ,
\label{eq:stokesPower}
\end{equation}
where $L_{\mathrm{eff}}={\int^L_0{P_{\mathrm S}(z) \, dz}}/{P_{\mathrm{S}}(L)}$ is the effective fiber length. It refers to the length of an amplifier with constant signal power $P_{\mathrm{S}}(L)$ along $z$ and longitudinally integrated power equal to $\int^L_0 {P_{\mathrm S}(z) \, dz}$. In an amplifier with $ P_{\mathrm S}(z) < P_{\mathrm{S}}(L)$, $L_{\mathrm {eff}}$ is shorter than the fiber length $L$. 

Our theory allows us to calculate the conventional SBS threshold based on the the condition that the steady-state Stokes power reflected exceeds a set fraction of $rP_s(L)$ of the signal power, (unfortunately this fraction varies widely in different works, from $0.1 \%$ and $10\%$, we chose 1\%). We use the excellent approximation that  the total reflected power is dominated by the mode with the highest Stokes gain, $P_{\mathrm B}^{(m)}(0)$.
With these assumptions, the amplified signal power at the SBS threshold is given by
\begin{equation}
    P_{\mathrm S}(L)= \frac{1}{L_{\mathrm {eff}}} \,  \frac{1}{\tilde{G}_{\mathrm B}^{(m)}} \, \ln \left[ \frac{r \, P_{\mathrm S}(L)}{P^{(m)}_{\mathrm B}(L)}\right] \, .
\end{equation}
Ignoring its weak (logarithmic) dependence on the output power level, the SBS threshold is inversely proportional to the effective fiber length $L_{\mathrm {eff}}$ as found experimentally in Fig.~2b. The threshold is also inversely proportional to the effective Brillouin gain coefficient $\tilde{G}_B^{(m)}$.  

Experimentally we measure the SBS instability threshold~\cite{harrison1990evidence, gaeta1991stochastic, lee2006chaotic, Agrawal2006,   melo2013destructive, panbiharwala2018experimental}, at which significant spikes appear in the time trace of reflected intensity.  Such spiking results from SBS and occurs at power level below the conventional SBS threshold. It prevents further power scaling as large Stokes pulses could damage the upstream lasers. Since the instability is gain-enhanced, it should depend on the average SBS gain calculated in our theory, and its threshold is expected to follow a similar scaling with $L_{\mathrm {eff}}$ as the conventional SBS threshold. As noted, this conjecture is confirmed in Fig.~2b of the main text. In order to compare to experiment, we assume that there is some reflectivity, $r_{sp}$, at which spiking arises and treat the logarithmic term in Eq. (8) as a single fitting parameter, which we set to agree with the lowest threshold, $L_{\mathrm {eff}} = 15.4$~m in Fig.~2b.

\begin{table}[t]
    \centering
    \begin{tabular}{|c|c|}
 \hline
 Parameter & Value \\ [0.5ex] 
 \hline\hline
 Fiber core diameter [\textmu m] & 42 \\ 
 \hline
 Fiber cladding diameter [\textmu m] & 570\\
 \hline
 Core numerical aperture & 0.1\\
 \hline
 Signal wavelength $\lambda_{\mathrm S}$ [nm] & 1064\\
 \hline
 Pump wavelength $\lambda_{\mathrm P}$ [nm] & 971 \\
 \hline
 Pump absorption cross-section  $\sigma^{\mathrm (a)}_{\mathrm P}$  [$\mathrm m^2$] & 1.2 $\times$ $10^{-24}$\\ 
 \hline
    Pump emission cross-section $\sigma^{\mathrm (e)}_{\mathrm P}$  [$\mathrm m^2$] & 1.02 $\times$ $10^{-24}$\\ 
 \hline
 Signal absorption cross-section $\sigma^{\mathrm (a)}_{\mathrm S}$  [$\mathrm m^2$] & 5.8 $\times$ $10^{-27}$\\ 
 \hline
 Signal emission cross-section $\sigma^{\mathrm (e)}_{\mathrm S}$  [$\mathrm m^2$] & 2.71 $\times$ $10^{-25}$\\ 
 \hline
 Yb concentration $\rho_{t}\: [\mathrm {ions} \, m^{-3}]$ & 6.5 $\times$ $10^{25}$\\
 \hline

    \end{tabular}
    \caption{Table of MMF amplifier parameters}
    \label{tab:B1}
\end{table}

	\end{appendices}
\end{document}